**Retracted Articles about COVID-19 Vaccines Enable Vaccine Misinformation on Twitter**

Rod Abhari
Esteban Villa-Turek
Nicholas Vincent
Henry Dambanemuya
Emőke-Ágnes Horvát

Northwestern University

*Research Article*

# Retracted Articles about COVID-19 Vaccines Enable Vaccine Misinformation on Twitter

*Retracted scientific articles about COVID-19 vaccines have proliferated false claims about vaccination harms and discouraged vaccine acceptance. Our study analyzed the topical content of 4,876 English language tweets about retracted COVID-19 vaccine research and found that 27.4% of tweets contained retraction-related misinformation. Misinformed tweets either ignored the retraction, or less commonly, politicized the retraction using conspiratorial rhetoric. To address this, Twitter and other social media platforms should expand their efforts to address retraction-related misinformation.*

## Research questions
- What types of misinformation are seen in tweets about retracted COVID-19 vaccine articles?
- How prevalent is retraction-related misinformation on Twitter?
- How broadly does retraction-related misinformation circulate through retweets?
- How might scientific and social media stakeholders address retraction-related misinformation?

## Essay summary
- Retracted COVID-19 vaccine articles were identified using a keyword search for 'vacc' in the Retraction Watch database of Retracted Coronavirus Articles up to June 2022. Tweets mentioning these articles were then identified using the Altmetric and Twitter API. All English language tweets and retweets posted after an article's retraction date were retained. This produced a corpus of 4,876 tweets containing 1,114 original tweets and 3,762 retweets. Each original tweet was hand coded for misinformation by three human coders.
- Just over 27% of tweets contained retraction-related misinformation. Of these, 85.7% of misinformed tweets were retraction-naïve tweets that shared the original article without criticizing its results, while the remaining 14.3% were retraction-politicized tweets that claimed the retraction was motivated by a hidden political agenda. (Findings 1 & 2).
- By connecting users that have retweeted a given tweet to the original tweet poster in a retweet network, we were able to analyze the circulation and visibility of misinformation. We found that attention to retracted COVID-19 vaccine articles took place in distinct retweet communities whose users tended to share the same type of information. While the largest retweet community almost exclusively shares reliable information about retracted COVID-19 vaccine articles, the other communities primarily shared misinformation (Finding 3).
- Efforts to address retraction-related misinformation by flagging retraction-mentioning content may be effective at counteracting retraction-naïve tweets. However, they are unlikely to effectively address retraction-politicized tweets, which can reframe content moderation as further proof of their conspiratorial premise.

## Implications

At the time of writing, roughly 270 research articles about COVID-19 have been retracted (Retraction Watch, 2020). Retractions are issued against academic research that contains serious research or publishing errors, including plagiarism and data fabrication (Barbour et al., 2009). Retracted COVID-19 research has jeopardized public health efforts in numerous ways, including by supporting untested treatments (Samaha et al., 2021), casting doubt on the effectiveness of face masks (Bae et al., 2020), and claiming unfounded harms from vaccines (Kafil et al., 2021). Once a retraction has occurred, notices of retraction can cause additional harm by damaging public trust in science. In the context of politicized scientific attitudes, prominent retractions may be interpreted as evidence of scientific incompetency or even conspiracy (Hall-Jamieson, 2021; Peterson et al., 2022). Retraction related misinformation can therefore result from both the publication of false information and from the subsequent undermining of trust in science.

One of the areas most impacted by retracted research is vaccine policy. Since the start of the pandemic, a number of articles erroneously claiming harms from COVID-19 vaccines have been retracted. These articles have likely fueled concerns about vaccine side-effects, which is the most cited reason for COVID-19 vaccine hesitancy and refusal (Warren et al., 2022; Yigit et al., 2021). One retracted vaccination paper claiming that vaccines would cause acute heart disease for 1 in every 1,000 vaccination cases overestimated the illness rate by 25-fold due to a calculation error–in actuality, there was a *decrease* in heart disease among the vaccinated (Kafil et al., 2021). By the time the article was retracted, however, it had received over 7,000 Twitter mentions and generated considerable negative press coverage against COVID-19 vaccines. The costs of vaccine misinformation are tangible: individuals who believed covid vaccine misinformation were less likely to take COVID-19 vaccines, and, in turn, were more likely to die due to COVID-19 complications (Albrecht, 2022).

These cases underscore the need to understand the circulation of retracted vaccine research during a global pandemic. This is especially true of social media mentions of retracted research. Although science communicated on social media can reach new audiences, the limited attention economy of social media can produce oversimplified or misleading interpretations of source material, making it more susceptible to information distortions (Caulfield et al., 2016; West & Bergstrom, 2021). In addition, polarized "echo chamber" environments within networked social media incentivize individuals to evaluate scientific information based on how well it conforms to their prior partisan beliefs (Del Vicario et al., 2016; Druckman & McGrath, 2019). Thus, while social media is not solely to blame for the 'infodemic' of COVID-19 misinformation, it has doubtlessly contributed to it.

*Types of retraction-related misinformation*
Two types of misinformation can arise from retracted vaccine tweets. First, misinformation can spread from tweets that present the original article as a scientific fact without acknowledging its retraction. As an unambiguous repudiation of an article's scientific legitimacy, retractions are one of the most important checks on scientific misinformation. However, this check can only exist when retractions are acknowledged. In academic publishing, authors are expected not to cite retracted papers for reasons unrelated to their retraction (Barbour et al., 2009). Whether this expectation exists on social media, however, is an open question. While recent research suggests that Twitter mentions of retracted research drop off after an article is retracted, the most likely reason is that the retraction comes after the initial attention to an article has been exhausted (Peng et al., 2022; Serghiou et al., 2021). When the retraction of a science article is not acknowledged, discredited scientific research may continue to propagate. Accordingly, retraction-naïve tweets directly undermine one of science's most essential checks against scientific misinformation.

Second, misinformation can spread from tweets that acknowledge a retraction, but frame it as being politically motivated. These tweets employ the rhetoric of science skepticism, an insidious form of scientific misinformation that does not explicitly deny the relevance of science, but instead seeks to diminish it by politicizing the motives of scientists (Oreskes & Conway, 2011). Utilizing science skepticism allows politicized actors to undermine mainstream ideology without committing to falsifiable positions, e.g., by 'just asking questions.' In the context of COVID-19, science skepticism has characterized much of the response to the pandemic among conservative figureheads (Hall-Jamieson, 2021). Science skepticism has also provided the breadcrumbs for the conspiracy-prone to piece together a sinister narrative in which scientists consort with "political elites" who direct the outcomes of their research. Although there is no evidence of this having occurred for any of the retracted COVID-19 vaccine articles discussed on social media, belief in conspiracy theories have nonetheless emerged as one of the key predictors of vaccine resistance (Romer & Hall-Jamieson, 2020, Romer & Hall-Jamieson, 2022).

*Solutions to retraction-related misinformation*
Although it once had one of the most robust policies for addressing COVID-19 misinformation, under its current CEO, Elon Musk, Twitter has quietly ended its policy (Lorenz, 2022). We nevertheless believe that it is important for social media platforms, including but not limited to Twitter, to (re)consider policies for addressing retraction-related misinformation, especially as it intersects with health misinformation. To this end, we discuss some approaches to addressing retraction-related misinformation that could be feasibly adopted by social media providers.

One possibility is to incorporate a retraction warning label, similar to how some social media give a content warning when COVID-19 misinformation is shared. The Zotero citation manager uses the Retraction Watch database of retracted articles to automatically flag retracted articles and warn users when they save a retracted article (Stillman, 2019). A similar feature could be added to social media platforms to alert posters when they share retracted articles and to alert subsequent readers that a post mentions a retracted article.

However, these efforts are unlikely to address misinformation produced by retraction-related tweets that politicize the retraction. Since the beginning of the pandemic, prominent media and political figures have exploited the inherent uncertainties of the scientific method to reduce support for COVID-19 vaccinations (Bolsen & Palm, 2022). These attacks have created a bulwark of resistance against efforts to correct vaccine-related misinformation through fact-checking and content warnings. Further, given that science skeptics use retraction as further evidence of a conspiracy to cover up the truth, heavy-handed moderation efforts may well backfire. A central tenet of conspiratorial thinking is that modern institutions share a common interest (Hofstadter, 1964). Thus, when politicized-retraction tweets are moderated, both science and social media can be targeted as co-conspirators. If retractions are interpreted as a form of censorship, moderation of retracted content may be seen as proof of this premise.

A more subtle solution is to reduce the visibility of content that contains retraction-related misinformation. Content reduction is a form of content moderation that deprioritizes problematic content in algorithmic recommendations (Gillespie, 2022). This is a policy that Elon Musk himself supports, having recently tweeted, "New Twitter policy is freedom of speech, but not freedom of reach." When combined with a reliable detection protocol, reduction may be an effective means of reducing exposure to retraction-related misinformation without generating accusations of political censorship.

Ultimately, retractions occupy a precarious position on social media sites that thrive on controversy and within a political climate hostile towards politically inconvenient research. Academic publishers may need to reconsider the utility of post-hoc corrections, including retractions and expressions of concern, and instead invest more resources into ensuring that dubious scientific research is not published in the first place. However, the most enduring solutions will require scientific and non-scientific stakeholders alike to address the deep chasms of polarization in the current political climate. The fate of scientific discourse may depend on it.

## Findings

*Finding 1: Misinformed original tweets were more frequent, but less retweeted, than other original tweets.*

We evaluated the information veracity of 4,876 English-language tweets mentioning at least one retracted COVID-19 vaccine article tweeted after the article's retraction. Within this corpus, 1,114 tweets were original tweets, and 3,762 tweets were retweets. Despite both being considered 'tweets,' original tweets contain a composed message, while retweets share the message of an original tweet without modification. Figure 1 shows that while 587 (52.7%) of original tweets contained misinformation, misinformed original tweets received only 749 (19.9%) total retweets. When original tweets and retweets are combined in Figure 2, misinformation is found in 27.3% of tweets. This indicates that although Twitter users composed more misinformed tweets than reliable tweets, accounting for retweets reduces the overall prevalence of misinformed tweets from roughly half of the dataset to a quarter of it.

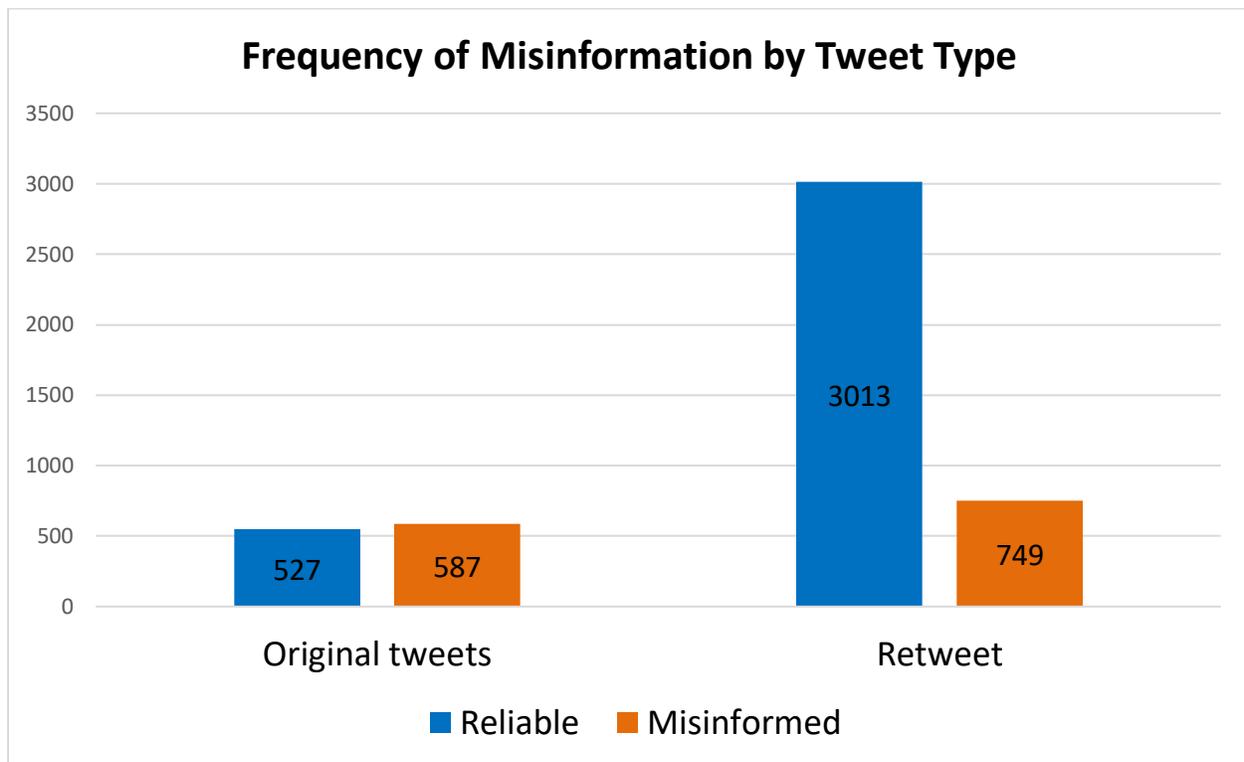

*Figure 1. Frequency of Misinformation by Tweet Type.*

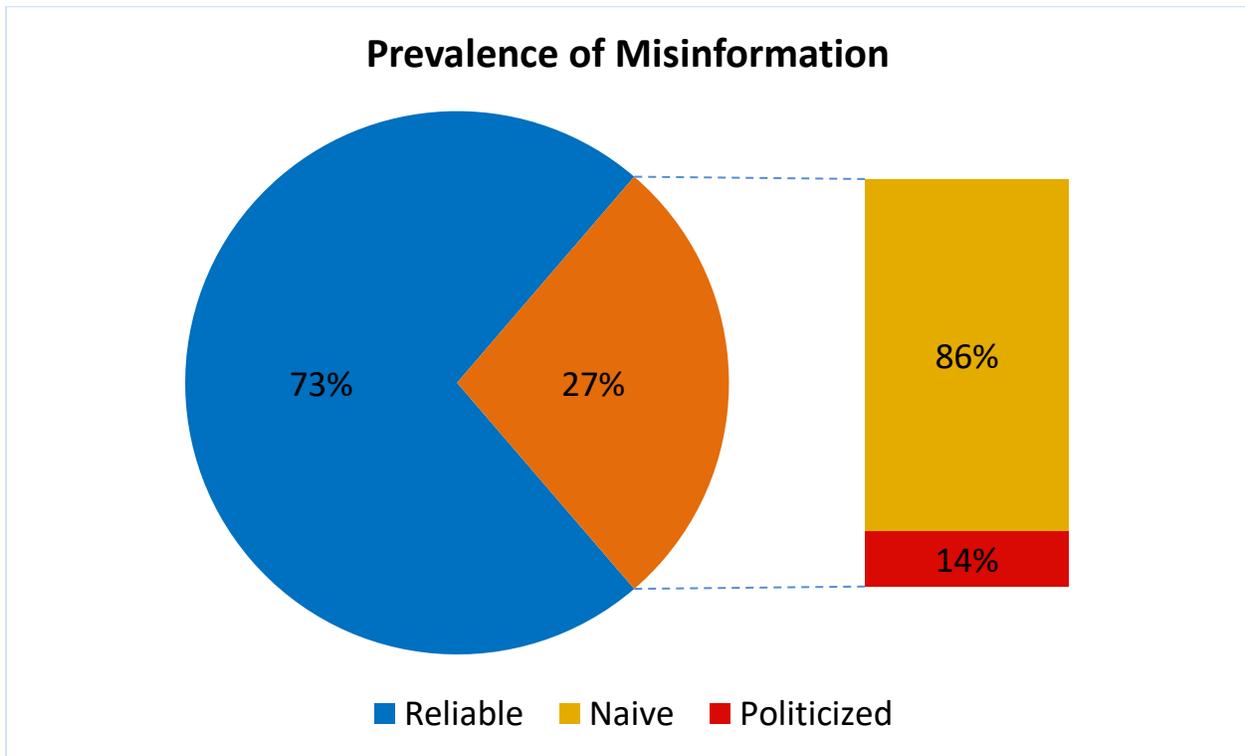

*Figure 2. Total Prevalence of Misinformation.*

*Finding 2: Among misinformed tweets, tweets that were naïve to the retraction were more prevalent than those that politicized it.*

Two types of retraction-related misinformation were found in the corpus. As shown in Figure 2, 85.7% of misinformed tweets were found to contain misinformation that resulted from the tweet failing to acknowledge the retraction which had already occurred, while the remaining 14.3% acknowledged the retraction, but instead framed it as motivated by political interests.

Retraction-naïve tweets misinform by failing to mention an article's retraction, providing the false impression that a retracted article's findings are scientifically sanctioned. A peculiarity of retraction-naïve tweets is they may be retweets of original posts about an article which occurred before its retraction. In Figure 3, for instance, the tweet highlighted in yellow was originally tweeted before the retraction but endured after retraction in the form of retweets. Because a tweet can be retweeted at any point, tweets that were originally posted before retraction can spread misinformation if they are retweeted after retraction, perhaps in spite of the intentions of the original poster. In these cases, our analysis only considers the post-retraction retweets to be retraction-naïve, not the original post.

| Retracted Article | Top Tweets | Stance |
|---|---|---|
| **Title**<br>"mRNA COVID-19 Vaccination and Development of CMR-confirmed Myopericarditis"<br><br>**Digital Object Identifier:**<br>10.1101/2021.09.13.21262182 | Canadian study that found an absurdly high myocarditis rate of 1 in 1000 among ppl getting mRNA vaccines was completely wrong, because study authors reported that 32,000 doses had been given, when in fact 800,000 had. They were only off by 25fold. | Retraction Aware |
| | Truth from Canada, up to date research on #vaccine and myocarditis<br>They are not safe for our children.<br>#usforthem<br>#childvaccines #Truth | Retraction Naive |
| | There is a corporation that is trawling the web for any doctors who are questioning their safety record.<br>And then targeting their employment, registration or manuscripts.<br>If you do nothing, you will be living under Mengele's regime. | Retraction Politicized |

*Figure 3. Reliable and Misinformed Tweet Examples.* Tweets were found in the corpus and assigned a stance by human coders.

Less prominent but more pernicious, retraction-politicized tweets dismiss the scientific credibility of the retraction by questioning its motives. As shown in Figure 3, retraction-politicized tweets generally frame the retraction of COVID-19 vaccine papers as a conspiracy to hide the underlying truth of vaccine harms. The tweet highlighted in red, for instance, imagines a corporation that uses retractions to censor whistleblowers within the scientific community. This plot is pure fiction, but it diverts discussion away from the scientific merits of the retracted article, and crucially, further entrenches scientific distrust within these communities.

*Finding 3. Retraction-related misinformation is concentrated in network communities where reliable tweets do not broadly circulate.*

In order to find the communities where retraction-related misinformation was most visible, we reconstructed the flow of information by connecting all users who posted at least one tweet from the corpus with the users who retweeted their tweet. This network is shown in its entirety in Figure 4.

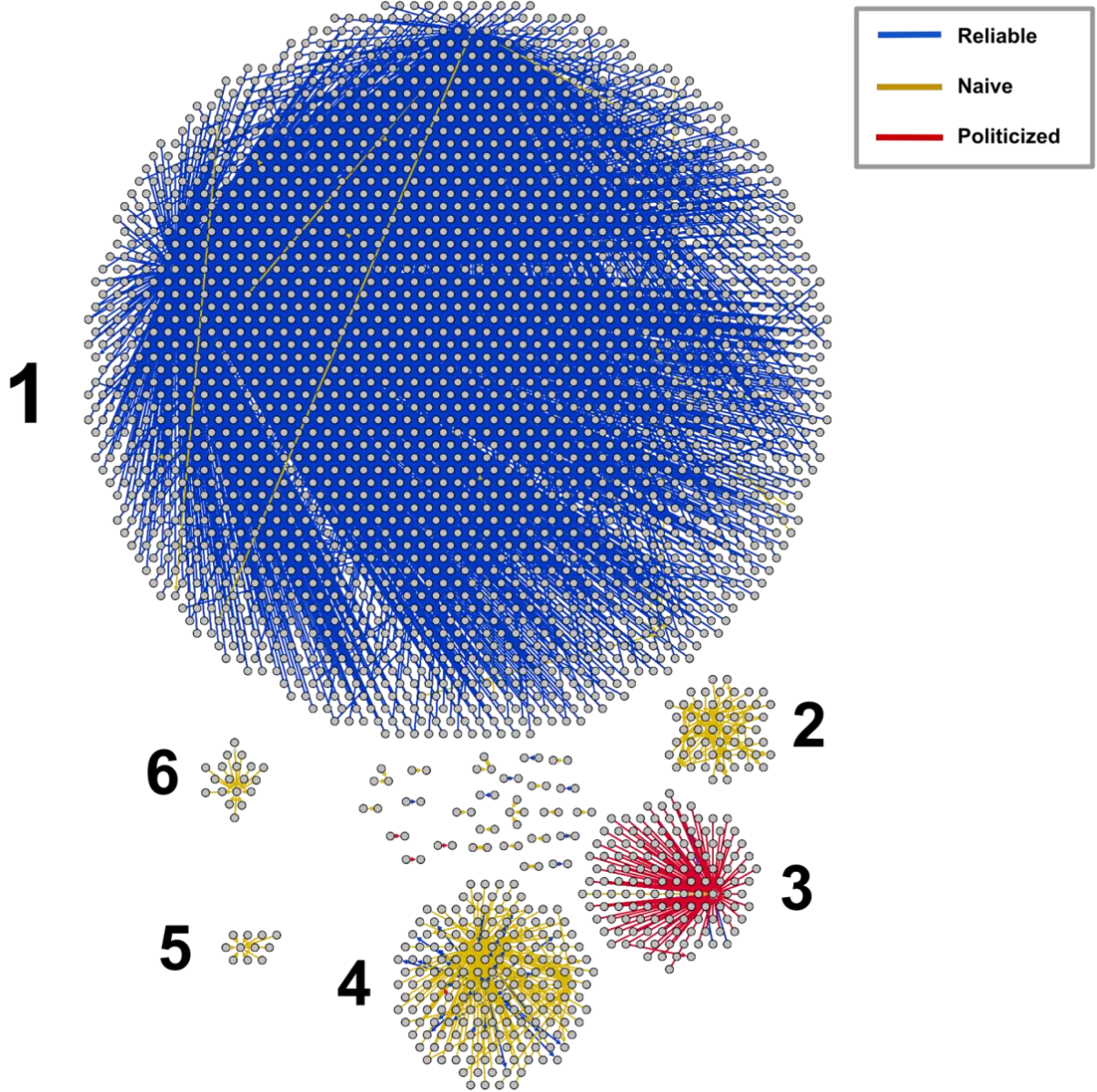

***Figure 4. Diffusion of Misinformed Tweets.*** *Colored lines connect nodes that have shared the same tweet. Numbered groups of nodes represent retweet communities detected by* Infomap. *Networks were visualized in Gephi using the* Circle Pack *Algorithm.*

We found that vaccine related information did not circulate evenly throughout the network. The bulk of reliable tweets were found in the largest community, community 1, with sparse attention in communities 3 and 4. Among misinformed tweets, naïve tweets circulated broadly in the smaller communities, while politicized tweets dominated community 3. Finally, there was a smaller cluster of users whose tweets were too sparsely retweeted to form a proper community.

These findings show how discussions of retracted COVID-19 articles occur in disconnected retweet communities where Twitter users interact selectively with particular kinds of retraction-related information. While users in the largest community avoid sharing retraction-related misinformation, there are also several communities where members exclusively retweet retraction-related misinformation. There are even clear demarcations in the type of misinformation shared, with naïve tweets occupying different network communities than politicized tweets. These findings demonstrate how users' exposure to misinformation is affected by the underlying community structure of Twitter interactions.

## Methods

Our methods consisted of collecting tweets about relevant articles, identifying distinct themes within the tweets, and reconstructing the retweet network to study the circulation of tweets. Below we describe the procedure for each method.

*Tweet collection*
First, we identified retracted COVID-19 articles about vaccines. To do so, we queried the Retraction Watch public database of retracted COVID-19 articles for all entries containing the keyword query for '-vacc,' identifying 18 retracted COVID-19 vaccine articles published between March 2020 and June 2022. For each article we retained the Digital Object Identifier (DOI) the official retraction date from the article webpage.

Second, we collected all relevant post-retraction Twitter mentions of each article. To do so, we used the API of "Altmetric.com," a company that tracks/indexes online mentions of academic research. By querying each DOI in the Altmetric API, we were able to gather Tweet and User IDs for all recorded tweets of the articles. Altogether, 9 of the 18 articles originally identified had publicly available tweets. We then queried each Tweet ID in the Twitter API to retrieve comprehensive information for each post, retaining the username, posted date, text, and response ids for each post. Because only post-retraction Tweet mentions are relevant to our analysis, we filtered out Tweets that were posted before the retraction date.

This process produced an initial corpus of 13,367 tweets of 9 articles. From the initial corpus, we retained all tweets where the entire text was in English. This produced our final corpus of 4,876 tweets of 9 articles, or 36.5% of the initial corpus. The final corpus contained 1,114 original tweets and 3,762 retweets.

*Thematic coding*
Next, we coded the text of all 1,114 original tweets using a method known as guided discourse analysis (Wetherell et al., 2001). Following this approach, the lead author reviewed the text of the tweets and identified two possible content characteristics. The two content characteristics were "whether a tweet mentions a retraction" and "whether a tweet ascribes a misleading or incorrect motivation for retraction." To facilitate analysis, the second content code was broken down into four mutually exclusive categories; namely, whether the tweet author ascribed the reason for retraction as "unstated or unclear," "scientific," "political," or "other." The coders were given instructions to consider political as a category that included "motivations to further the economic or political interests of non-scientific actors." The "political" and "other" motivations for retraction were *a priori* determined to be misinformation categories. This is because none of the retractions of the articles in the corpus could reasonably be interpreted as having non-scientific motivations.

Two coders were then recruited for the purpose of assigning content codes to each tweet. Cohen's kappa was measured for each item to assess intercoder agreement. The authors received high consensus on their codes, with agreement ranging from "substantial agreement" (κ = 0.728) to "almost perfect agreement" (κ = 0.89). The full values are included in the Appendix. For the 142 tweets with disagreement, an author acted as tiebreaker and resolved the dispute. Finally, it is worth noting that the 'other' category did not receive any tweet instances in the final corpus. Thus, we were able to drop the category. This also confirmed the utility of the 'political' category in its coverage of tweets that ascribed non-scientific reasons for retraction.

*Network Reconstruction*
We reconstructed the network of retweets by using the "author id" field returned by the Twitter API to identify the source author of tweets, and then connecting authors of retweets to the authors of the original tweet using the "retweet id" fields. We used a directed multigraph network model to allow for multiple tweets between authors. For each tweet, we included the tweet code assigned by the coders to ensure the possibility of parallel retweet relationships between reliable, naïve, and politicized tweets separately.

The network was visualized using *Gephi*. Using *Gephi*, we filtered out node isolates, or users that did not tweet any retweeted tweets. We then used a community detection algorithm, *Infomap*, to assign nodes a module id. While there are numerous ways to measure communities, *Infomap* has been identified as a particularly effective way to delineate communities within communication networks (Rosvall & Bergstrom, 2008). Finally, nodes were automatically grouped into their *Infomap* community using the *Circle Pack Layout* plugin.

**Acknowledgements (optional)**
The authors are grateful for the assistance of Retraction Watch in providing a database of retracted articles. Additionally, they thank Altmetrics for providing access to their API which was used to collect data on Altmetric counts for articles. The authors would also like to acknowledge the extensive work of Anna Cable for their help in coding Tweets.

**Funding**

**Competing interests**
There are no competing interests to report.

**Ethics**
The research did not use data from vulnerable or protected human subjects. When using the Twitter API, only tweets relevant to the discussion of article retractions were retained. Tweets were only collected from public Twitter accounts.

# Appendix: Tweet Coding

*Table 1.* Inter-rater Reliability for Coded Tweets.

| Category | Reliability | Coder A | Coder B | Errors |
|---|---|---|---|---|
| Retraction not mentioned | 0.890 | 569 | 554 | 15 |
| Retraction mentioned, reason not provided | 0.728 | 313 | 309 | 4 |
| Retraction mentioned, scientific reason provided | 0.754 | 187 | 209 | 22 |
| Retraction mentioned, political reason provided | 0.749 | 44 | 43 | 1 |
| Retraction mentioned, other reason provided | 0 | 1 | 0 | 1 |

*Note: Inter-rater reliability determined by calculating Cohen's Kappa for each item.*